\documentclass[12pt]{article}

\usepackage{newtxtext,newtxmath}
\usepackage{graphicx}

\usepackage[letterpaper,margin=1in]{geometry}

\renewenvironment{abstract}
	{\quotation}
	{\endquotation}

\date{}

\makeatletter
\renewcommand{\fnum@figure}{\textbf{Figure \thefigure}}
\renewcommand{\fnum@table}{\textbf{Table \thetable}}
\makeatother

\usepackage{url}

\def\scititle{
	Topological photonic crystal fiber
}
\title{\bfseries \boldmath \scititle}

\author{
	Bofeng Zhu$^{1,6\dagger}$,
	Kevin Hean$^{2\dagger}$,
	Stephan Wong$^{3}$,
	Yuxi Wang$^{2}$,\and
	Rimi Banerjee$^{1,6}$,
	Haoran Xue$^{4}$,
	Qiang Wang$^{5}$,
	Alexander Cerjan$^{3}$,\and
	Qi Jie Wang$^{1,2,6\ast}$,
	Wonkeun Chang$^{2\ast}$,
	Y. D. Chong$^{1,6\ast}$,\and
	\small$^{1}$ School of Physical and Mathematical Sciences, Nanyang Technological University, Singapore \& 637371, Singapore.\and
	\small$^{2}$ School of Electrical and Electronic Engineering, Nanyang Technological University, Singapore \& 639798, Singapore.\and
	\small$^{3}$ Center for Integrated Nanotechnologies, Sandia National Laboratories, Albuquerque, New Mexico \& 87185, USA.\and
	\small$^{4}$ Department of Physics, The Chinese University of Hong Kong, Shatin, Hong Kong SAR \& China.\and
	\small$^{5}$ School of Physics, Nanjing University, Nanjing \& 210093, China.\and
	\small$^{6}$ Centre for Disruptive Photonic Technologies, Nanyang Technological University, Singapore \& 637371, Singapore.\and
	\small$^\ast$Corresponding author. Email: qjwang@ntu.edu.sg, wonkeun.chang@ntu.edu.sg, yidong@ntu.edu.sg\and
	\small$^\dagger$These authors contributed equally to this work.
}

\begin{document} 

% Insert the title and author list
\maketitle

% Abstract, in bold
% There are strict length limits, and not all formats have abstracts.
% Consult the journal instructions to authors for details.
% Do not cite any references in the abstract.
\begin{abstract} \bfseries \boldmath
Photonic crystal fibers (PCFs) provide an exceptionally versatile platform for various applications, thanks to the flexibility with which light-guiding can be customized by modifying the fiber geometry. Here, we realize a PCF with guided modes produced by photonic bandstructure topology rather than conventional mode-trapping mechanisms. The design, which is compatible with the stack-and-draw fabrication process, consists of a cross sectional photonic topological crystalline insulator with a disclination.  A bulk-defect correspondence produces degenerate topological modes, lying below the cladding light line.  We use various theoretical methods to confirm their topological origins, including a spectral localizer that makes minimal assumptions about the bandstructure.  Our experiments on the fabricated topological fiber show it transmitting visible to near-infrared light with low losses of 10--20 dB/km, which do not increase significantly when the fiber is bent.  A comparable solid-core PCF of conventional design exhibits substantially higher bending losses.  Optical fibers based on topological modes thus hold promise for improved performance and versatile functionalities.
\end{abstract}

\noindent
\textbf{Teaser sentence} (111 characters): This work demonstrates a photonic crystal fibre using topological protection from its nontrivial bandstructure.

\section*{Introduction}

\noindent
PCFs \cite{Joannopoulos1997, Knight2003, MarkosReview2017, Fokoua2023, Birks1997, Temelkuran2002, Charlene2003, Couny2007, Beravat2016} are a subset of the broader class of photonic crystals: structures that use wavelength-scale modulations to manipulate light \cite{Joannopoulos1997, Joannopoulos2008}.  Although photonic crystals have been used in high-performance lasers \cite{Noda2017} and solar cells \cite{Liu2019}, arguably their most important applications are in PCFs, including  high-power light delivery \cite{Temelkuran2002}, supercontinuum light generation \cite{SylvestreReview2021}, and sensing \cite{ChaudharyReview2022}.  In recent years, a new approach to designing photonic crystals, called topological photonics, has emerged \cite{OzawaReview2019, TangReview2022, Vaidya2023}.  This involves engineering photonic bandstructures similar to those found in topological phases of condensed matter, thereby giving rise to distinctive photonic modes that owe their existence to topological ``correspondence principles'' rather than conventional light-trapping mechanisms \cite{Benalcazar2019, Li2020, Peterson2021, Liu2021}.  Aside from providing new avenues for fundamental research into band topology, topological photonics holds promise for device applications due to the robustness of topological modes against certain forms of disorder.  There is a substantial amount of ongoing research on topological waveguides \cite{Barik2016, Ma2016} and resonators \cite{Bahari2017, ZengYQ2020, Dikopoltsev2021}, mostly based on the photonic crystal slab geometry, which is well-suited to photonic bandstructure engineering due to the availability of powerful fabrication techniques like photolithography.  Bringing topological photonics into PCFs, however, has proven more challenging.

Previous theoretical proposals \cite{Lin_fiber2020, Pilozzi2020, Makwana_fiber2020, Gong_fibre2021, Huang_fiber2023} for topological photonic crystal fibers have been hampered by incompatibility with existing fiber fabrication methods, including the lack of mechanical stability in preform stacking, reliance on delicate structural features or precise index modulation in glass, and other issues.  Many other designs for implementing topological photonics in PhC slabs are difficult to adapt to PCFs for similar reasons.  Recently, researchers have developed a multicore PCF whose cores are placed in a Su-Schrieffer-Heeger configuration, a one-dimensional lattice with topological end-states \cite{Roberts2022}, but that design is based on the arrangement of the waveguiding cores, not the topological properties of the underlying PCF bandstructure.  

Here, we design and experimentally implement a topological photonic crystal fiber (TPCF) that guides light via robust topological modes based on defect states in topological crystalline insulators (TCIs).  Although studies of topological bandstructures usually focus on boundary states created by the bulk-boundary correspondence principle \cite{OzawaReview2019}, lattice defects such as disclinations \cite{Mermin1979} can also host localized topological states due to the related bulk-defect correspondence principle \cite{Lin2023NRP}, as shown in recent experiments \cite{Liu2021, Peterson2021, Hwang2024, Hu2024}.  (Photonic defect modes based on other topological principles have also been demonstrated \cite{Wang2020, Gao2020, Yang2022, Han2023}.)  In particular, TCIs---structures with nontrivial band topology sustained by lattice symmetries \cite{Benalcazar2019, Li2020, Vaidya2023}---can host topological disclination states associated with fractionalized spectral charge, as shown recently in a pair of photonics-based experiments \cite{Peterson2021, Liu2021}.  Taking a similar approach, we design a TPCF whose two-dimensional (2D) cross section is a photonic TCI with a central disclination.  The overall structure, consisting of glass capillaries and rods of different radii, is compatible with the standard stack-and-draw method for fabricating PCFs.  The TCI hosts disclination states localized around a central air hole; when extruded along the fiber's $z$ axis, these form a set of ten waveguide modes that we call guided topological defect modes (GTDMs).  To show that the GTDMs originate from nontrivial photonic band topology, we first establish that the underlying bulk TCI is topologically nontrivial and that the disclination traps fractional charges \cite{Peterson2021, Liu2021}; then, we employ a recently-developed ``spectral localizer'' framework \cite{Loring2015, Cerjan2022, JMP2023, Cerjan2024, Alex_APL2024, Wong2024, Garcia2024} to specifically identify the GTDMs as topological states.

These GTDMs possess an unexpected feature that is advantageous for waveguiding: despite originating from topological gaps in the bulk bandstructure, they do not reside in those gaps, nor in the bulk bands.  Instead they lie below the lowest bulk band, i.e., below the cladding light line. This property, which is not disallowed by theoretical principles \cite{Benalcazar2019, Li2020}, serves to inhibit cross-talk with the bulk states in a manner analogous to solid-core PCFs \cite{Birks1997}.
%% \textcolor{magenta}{Compared with the disclination defect modes in the photonic crystal slabs \cite{Peterson2021,Liu2021}, the GTDMs are spectrally separated from the bulk bands and do not require local lattice modifications, such as on-site detuning \cite{Peterson2021} or modified boundary conditions \cite{Liu2021}.}
It allows for robust, broad-band waveguiding without a complete 3D bandgap of the sort employed in earlier demonstrations of guided defect modes (which used designs that are hard to implement in a PCF) \cite{Noh2020, Qiang2021, Hu2024}.  We experimentally measure the TPCF's transmission loss to be 10--20~$\textrm{dB/km}$ across much of the visible to near-infrared range.  When the TPCF is strongly bent (2 loops of radius 1 cm), the measured output power decreases by less than 5 dBm throughout the operating frequency range, whereas a conventional solid-core PCF fabricated with the same equipment and facilities experiences a decrease of up to 25 dBm.  The GTDMs also have an interesting structure that can be exploited for spatial and/or polarization multiplexing.  In the future, it will be interesting to see how such TPCFs, after appropriate perfomance optimizations, compare to state-of-the-art PCFs based on conventional design principles.  Our work also points to optical fibers as an important platform for future work on topological photonics, with many unexplored application possibilities.

\section*{Results}

\subsection*{Fiber design and implementation}
\label{secB}

\begin{figure}
  \centering
  \includegraphics[width=1\textwidth]{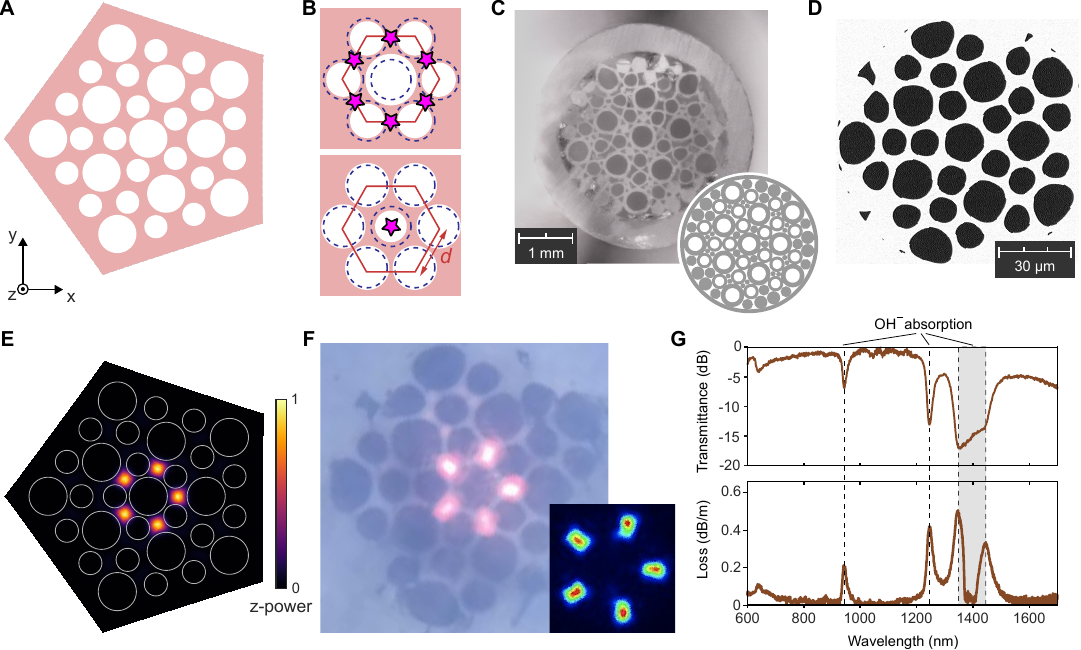}
  \caption{\textbf{Topological photonic crystal fiber (TPCF) implementation.} (\textbf{A}) Schematic of cross sectional photonic structure, comprising a topological crystalline insulator (TCI) with a disclination. (\textbf{B}) Unit cell (red lines) for the disclination-free TCI, which has $C_{6v}$ symmetry.  Starting from a triangular lattice of equal air holes (dashed circles), the hole radii are alternately increased and decreased.  For two different choices of modulation, the Wannier centers (magenta stars) are located at the unit cell's sides (upper plot) or center (lower plot). (\textbf{C}) Photograph of the drawn fiber cane.  Inset: stacking arrangment of the preform. (\textbf{D}) Scanning electron microscope image of the TPCF's end face. (\textbf{E}) Calculated intensity profile (power flow in the $z$ direction) for a disclination state of the structure from (\textbf{A}) at $k_z d/2\pi = 2$. The air holes are indicated by white circles.  (\textbf{F}) Optical microscope photograph of a $100$~$\textrm{m}$ TPCF, with a supercontinuum light source at the opposite end. Inset: infrared camera image of the same.  (\textbf{G}) Measured transmittance (upper plot) for a $67$~$\textrm{m}$ TPCF and transmission loss (lower plot) using the same source.}
  \label{fig:TPCF}
\end{figure}

The TPCF is translationally invariant along the fiber axis (denoted by $z$), with a pattern of air holes in the transverse ($x$-$y$) plane (Fig.~\ref{fig:TPCF}A).  This cross sectional pattern forms a 2D photonic crystal with a lattice defect \cite{Mermin1979, Lin2023NRP}. Specifically, it is a photonic TCI of the Wu-Hu type \cite{Huang_fiber2023, Wu2015, Barik2016}, containing a disclination that hosts topological disclination states \cite{Peterson2021, Liu2021, Lin2023NRP} to be used for waveguiding.

%\begin{figure}
%  \centering
%  \includegraphics[width=1\textwidth]{Fig1_v2}
%  \caption{\textbf{Topological photonic crystal fiber (TPCF) implementation.} (\textbf{A}) Schematic of cross sectional photonic structure, comprising a topological crystalline insulator (TCI) with a disclination. (\textbf{B}) Unit cell (red lines) for the disclination-free TCI, which has $C_{6v}$ symmetry.  Starting from a triangular lattice of equal air holes (dashed circles), the hole radii are alternately increased and decreased.  For two different choices of modulation, the Wannier centers (magenta stars) are located at the unit cell's sides (upper plot) or center (lower plot). (\textbf{C}) Photograph of the drawn fiber cane.  Inset: stacking arrangment of the preform. (\textbf{D}) Scanning electron microscope image of the TPCF's end face. (\textbf{E}) Calculated intensity profile (power flow in the $z$ direction) for a disclination state of the structure from (\textbf{A}) at $k_z d/2\pi = 2$. The air holes are indicated by white circles.  (\textbf{F}) Optical microscope photograph of a $100$~$\textrm{m}$ TPCF, with a supercontinuum light source at the opposite end. Inset: infrared camera image of the same.  (\textbf{G}) Measured transmittance (upper plot) for a $67$~$\textrm{m}$ TPCF and transmission loss (lower plot) using the same source.}
%  \label{fig:TPCF}
%\end{figure}

In the absence of the disclination, the cross sectional structure has a $C_{6v}$ point group symmetry, with each unit cell containing two air holes (upper plot of Fig.~\ref{fig:TPCF}B).  We introduce a disclination of Frank angle $-\pi/3$ into the lattice using a ``cut-and-glue'' process \cite{Chaikin2000, Li2020} (see the Supplementary Materials, section S1 and Fig.~S1). This yields the five-fold rotationally symmetric structure of Fig.~\ref{fig:TPCF}A, in which the hole radii are $R_1 = 0.57d$ and $R_2 = 0.35d$, where $d$ is the nearest-neighbour center-to-center distance in the original periodic lattice (see Fig.~\ref{fig:TPCF}B).

By design, such a structure is compatible with the stack-and-draw process \cite{Knight2003, MarkosReview2017} for fabricating PCFs.  The air holes are formed from glass capillaries of two different radii, stacked within a glass jacket with major gaps filled by additional solid glass rods, as shown in the inset of Fig.~\ref{fig:TPCF}C.  This stacking arrangement is derived from the configuration of Fig.~\ref{fig:TPCF}A by a jam-packing procedure. The preform is drawn into a fiber cane (Fig.~\ref{fig:TPCF}C), which is then further drawn into a fiber of diameter 310~$\mu\textrm{m}$.  A scanning electron microscope image of the TPCF's end face is shown in Fig.~\ref{fig:TPCF}D, showing that the drawing process has filled in most of the interstitial air holes, while also slightly deforming the main air holes.  Further details about the fabrication procedure are given in the Supplementary Materials, section S1.

The key operating principle for the TPCF is that disclinations in TCIs can bind fractional spectral charges, giving rise to localized disclination states \cite{Peterson2021, Liu2021, Lin2023NRP}.  Such states are tied to the TCI topology via a bulk-defect correspondence, similar to the bulk-boundary correspondence governing topological corner states in disclination-free TCIs \cite{Benalcazar2019, Li2020} (in some circumstances, disclination states can even act as probes for band topological features that boundary probes cannot pick up).  Whereas previous experimental realizations of topological disclination states \cite{Peterson2021, Liu2021} have been based on structures that map closely to theoretical tight-binding models, our TPCF, like other photonic crystals, has no direct tight-binding analogue.  Nonetheless, its photonic bandstructure can be shown to have nontrivial topology giving rise to disclination states.  The Wannier centers are located on the sides of the periodic structure's unit cell, as shown in the upper plot of Fig.~\ref{fig:TPCF}B, implying that adding a disclination binds fractional charge \cite{Benalcazar2019, Peterson2021, Liu2021}. Consistent with this prediction, numerical simulations of Maxwell's equations on the structure of Fig.~\ref{fig:TPCF}A (see the Supplementary Materials, section S2) reveal the existence of guided modes that are strongly localized to the center of the sample (Fig.~\ref{fig:TPCF}E).

When light is coupled into the fabricated TPCF, we observe a spatially localized output profile at the end face (Fig.~\ref{fig:TPCF}F). These optical microscope and infrared camera images are taken using a $100$~$\textrm{m}$ long TPCF with a supercontinuum laser source coupled to the opposite end.  Further details about the experimental setup are given in Materials and Methods.  The light is concentrated at five high-index (glass) regions placed symmetrically around the central air hole, closely matching the prediction of Fig.~\ref{fig:TPCF}E.

We will further investigate the disclination states, and their relationship to TCI topology, in the next section.  It can be noted that reversing the large and small air holes yields a structure with trivial TCI topology, for which the Wannier centers lie at the centers of the unit cells (lower plot of Fig.~\ref{fig:TPCF}B).  In that case, introducing a disclination will induce neither charge fractionalization nor defect states (see the Supplementary Materials, section S3).

The measured transmittance spectrum of the TPCF (Fig.~\ref{fig:TPCF}G, upper plot) shows that it can operate across a substantial wavelength band of around $700$--$1200$~$\textrm{nm}$ and $1500$--$1650$~$\textrm{nm}$.  %The spectral peak at $1060\,\textrm{nm}$ is an inherent property of the supercontinuum source, while the dips are due to hydroxyl ion absorption \cite{Weik2017}.  
We also determine the loss spectrum (Fig.~\ref{fig:TPCF}G, lower plot) by comparing the outputs for TPCFs of length $67$~$\textrm{m}$ and $15$~$\textrm{m}$, thus normalising away the source spectrum (see Materials and Methods).  We find an average loss of around $10$~$\textrm{dB/km}$ over $1000$--$1200$~$\textrm{nm}$ and around $20$~$\textrm{dB/km}$ over $1530$--$1625$~$\textrm{nm}$.  Although we have not yet performed rigorous optimization of the TPCF to minimise transmission losses, the current performance is already close to the level in commercial solid-core PCFs of similar core size (e.g., around $8$~$\textrm{dB/km}$ at 1064 nm for Thorlabs LMA-25, which has a core size of 25~$\mu\textrm{m}$).

%% \textcolor{cyan}{[Input from Wonkeun] I would not try to compare the loss in the TPCF to that of hollow-core fibers, as they are two totally different classes of fibers. Comparing TPCF with solid-core PCFs is more meaningful, considering that in both TPCF and solid-core PCF, the light is guided in the silica glass, and there are microstructures that support the guidance in the solid glass. I think we can simply state our measured loss values and mention that we have not yet attempted to optimise our TPCF to reduce losses. Therefore, we should not not make any loss comparisons to either solid- or hollow-core fibers. If you think a comparison is necessary, I suggest we use the commercial PCF transmission loss value. Commercial photonic crystal fibers can have transmission loss as low as a few dB/km.}

% Meanwhile, two of the five nodes are excited and the mode profiles remain nearly unchanged }), which means the guided modes are fundamental mode (the mode with the largest $k_z$ for any given frequency) across a wide frequency range.

\subsection*{Guided topological defect modes}
\label{secC}

Like most other PCFs, the TPCF, considered as a three-dimensional (3D) structure, lacks a complete photonic bandgap.  Using its measured cross sectional structure (Fig.~\ref{fig:TPCF}D), we calculate the transverse eigenmodes at each axial wavenumber $k_z$, obtaining the band diagram shown in Fig.~\ref{fig:band_indicators_charges}A (for details about the eigenmode calculations, see the Supplementary Materials, section S2).  There are numerous bulk modes (pink-shaded regions), and a small number of disclination states that are strongly localized in the cross sectional plane (red-and-blue lines), which we call guided topological defect modes (GTDMs).  Over some ranges of $k_z$ on the order of $2\pi/d$, there are gaps in the transverse spectrum, but the GTDMs do not lie within them.  Although the bulk-defect correspondence predicts the disclination states, it does not require them to lie within a gap and there is no chiral-like symmetry pinning their frequencies \cite{Noh2020,Vaidya2023}.  Absent fine-tuning, disclination states in previous models have shown a similar tendency to migrate into the bulk state continuum \cite{Peterson2021, Liu2021}.

The eigenmode calculations reveal a total of ten GTDMs. With increasing $k_z$, they dive below the ``cladding light line'' defined by the lowest-frequency bulk fiber modes.  This feature is reminiscent of the ``scalar limit'' of guided modes in solid-core PCFs \cite{Birks1997}, and protects the GTDMs from coupling to bulk modes, similar to being in a bandgap.  In the large-$k_z$ regime, the GTDMs become effectively degenerate in frequency, as shown on the right side of Fig.~\ref{fig:band_indicators_charges}A; for reference, $20 \lesssim k_z d/2\pi \lesssim 30$ corresponds to the 730--1100~$\textrm{nm}$ operating regime of our experiments.  The GTDMs form two groups of five modes each; one group has in-plane electric fields polarized in the radial direction, while the other is azimuthally polarized (Fig.~\ref{fig:band_indicators_charges}B).  Their intensities are strongly localized to five high-index regions surrounding the central air hole, similar to the previous ideal case (Fig.~\ref{fig:TPCF}E).  Moreover, with increasing $k_z$, they exhibit rising in-plane quality ($Q$) factors and decreasing mode area (Fig.~\ref{fig:band_indicators_charges}C and D).

The location of the GTDMs below the cladding light also ensures that they have the lowest losses among all the guided modes of the TPCF.  As a consequence, when modes are launched in the fiber through standard butt-in coupling (as discussed in the next section), the other modes will be rapidly damped with propagation distance, so that only the GTDMs remain non-negligible.  This desirable feature removes the need to couple to the GTDMs via precise focusing of the input light.

%% However, it should be noted that the GTDMS are \textit{not} necessarily below the light-line. Indeed, as shown in the eigenmodes spectrum of deformed lattices (i) and (ii) in Fig.\textbf{\ref{fig:band_indicators_charges}b}, the GTDMs are found to be above light-line and embedded in the bulk bands, which are hardly to become guiding modes in fibers. However, after perturbing the positions of smaller airholes (see the lowerleft inset of Fig.\textbf{\ref{fig:band_indicators_charges}b}), the GTDMs are pushed below the light-line and become the fundamental guided modes of TPCF (the mode with the smallest $\omega$ for any given $k_z$.). Note that certain local deformations have been applied in $C_3$- and $C_5$-symmetric lattices so that the TDMs are in-gap \cite{Peterson2021}.

%As is known, conventional SC-PCF can remain single-mode operation regardless of the wavelength (the so-called \textit{endlessly single-mode}) \cite{Birks1997}. However, higher orders of guided modes are able to remain confined in the core if the air hole radius and lattice constant satisfies $r/d > 0.2$ \cite{MarkosReview2017}. In realistic case with fabrication-induced imperfections, there would always be one of the ten GTDMs to be less lossy than others. The TPCF can remain in \textit{endlessly single-mode} with sufficient long length even if all the five nodes are excited at the same time (see \textbf{Supplementary Movie S1 and S2 }) or only one node is directly excited (see \textbf{Supplementary Movie S3}).

To understand the topological origins of the GTDMs, we return to the underlying 2D TCI (Fig.~\ref{fig:TPCF}B, upper plot), which has undeformed circular air holes and is spatially periodic (disclination-free).  Its bands have zero Chern numbers, even for $k_z \ne 0$, and are thus Wannier representable \cite{Li2020}.   We plot the bulk spectrum of this periodic structure for $k_zd/2\pi = 2$ (Fig.~\ref{fig:band_indicators_charges}E, left panel); a low $k_z$ is chosen so that the bands are more easily distinguishable.  From this, we observe that the two lowest sets of bands, denoted by $\#(1,2,3)$ and $\#(4,5,6)$, are separated from the others by gaps.  We then analyze these bands using established methods for characterizing 2D TCIs \cite{Benalcazar2019, Li2020, Liu2021}.  By examining the phase profiles of their Bloch wavefunctions at high-symmetry momentum points, we are able to determine that both sets of bands have symmetry indicators $(\chi_M,\chi_K)=(2,0)$ \cite{Benalcazar2019, Li2020, Liu2021}.  Then, by varying the base point of a Wilson loop, we observe nontrivial Berry phases around $\pm \pi$ (Fig.~\ref{fig:band_indicators_charges}E, right panel), which corresponds to the Wannier centers being localized to the sides of the unit cell (Fig.~\ref{fig:TPCF}B, upper plot) \cite{Liu2021}.  For more details about the characterization procedure, see the Supplementary Materials, section S3 and Fig.~S2. The results of the analysis imply that each set of bands binds fractional spectral charge, which is a necessary condition for localized topological states to emerge \cite{Peterson2021}. To verify this, we turn to the preform hole profile (i.e., a finite structure containing a disclination), identify the GTDMs (Fig.~\ref{fig:band_indicators_charges}F, left panel), and calculate their spectral charges. For each polarization, we indeed find a charge of $\approx 0.5$ in each of these five areas (Fig.~\ref{fig:band_indicators_charges}F, right panel), in agreement with the TCI symmetry indicators. Note that an analogous tight-binding model has five disclination states arising from three bands \cite{Liu2021}, so the existence of ten GTDMs can be interpreted as a doubling due to the polarization degree of freedom.  The above properties also hold at the larger $k_z$ values where the TPCF operates.

To further confirm that the GTDMs are topological modes, we employ an independent characterization framework called the spectral localizer~\cite{Loring2015}.
By combining information about a system's position operators and Hamiltonian, the spectral localizer characterizes the system's topology directly in real-space.  This approach complements the spectral charge analysis since it does not refer to a periodic precursor lattice, and can accommodate the breaking of the TCI's protecting symmetries by the disclination and fabrication-induced deformations.  In the same spirit as topological band theory, the spectral localizer uses homotopy arguments to characterize a structure, e.g.~by assessing through a topological invariant whether its Hamiltonian can be continued to a trivial insulator.  In this case, the characterization is performed via (i) an index $\zeta_{\omega}^{\mathcal{S}}$ and (ii) a local gap measure $\mu_\omega$.  At each frequency $\omega$, $\zeta_{\omega}^{\mathcal{S}}$ classifies what kind of atomic limit the system is continuable to while preserving a local spectral gap and a stated symmetry $\mathcal{S}$; we choose $\mathcal{S}$ to be a global mirror symmetry $y \rightarrow -y$ (the $C_{6v}$ symmetry of the precursor lattice is not usable as it is broken by the disclination).  The values of $\zeta_\omega^\mathcal{S}$ can be integers or half-integers, which correspond to distinct topological crystalline phases that cannot be continuously deformed into each other without breaking $\mathcal{S}$ and/or closing the local gap; changes in $\zeta_\omega^\mathcal{S}$ are quantized to integers and correspond to the number of topological states at frequency $\omega$~\cite{Cerjan2024}.  Meanwhile, $\mu_\omega$ quantifies the degree of topological protection, in the sense that a topological state at $\omega$ is robust against perturbations $\delta H$ for which $\Vert \delta H \Vert < \mu_\omega$, where $\Vert \cdot \Vert$ denotes the largest singular value \cite{Alex_APL2024}.  For details, see the Supplementary Materials, section S4.  

The spectral localizer analysis reveals that the 10 GTDMs are all associated with jumps in $\zeta_{\omega}^{\mathcal{S}}$ (Fig.~\ref{fig:band_indicators_charges}G, left and middle panels), alongside nonzero $\mu_\omega$ at intermediate frequencies (Fig.~\ref{fig:band_indicators_charges}G, right panel).  These results confirm that the GTDMs are robust topological modes.  Furthermore, by extending the spectral localizer analysis using a generalized local gap measure \cite{Garcia2024}, we are able to show that the results remain valid even if the TPCF is bent along an arbitrary direction (which deform the structure in a manner that need not preserve the mirror symmetry $\mathcal{S}$, as explained in the next section); for details, see the Supplementary Materials, section S5.

\subsection*{Characterization of fiber properties}
\label{secD}

Having established the existence of GTDMs in TPCFs and their topological origins, we show that their properties are well-suited for waveguiding applications.

First, we verify that the theoretically-predicted features of the GTDMs, including their aforementioned degeneracy, polarization, and spatial structure, are preserved during actual light transmission through the TPCF.  Using the setup shown in Fig.~\ref{fig:experimental_data}A, we couple linearly polarized light at a single wavelength (1070~nm) into a 0.5~m long TPCF.  By adjusting the position of the beam spot on the input face of the TPCF, we find that the intensity profile at the end face can be strongly concentrated onto each of the five high-index regions around the central air hole (an example is shown in the inset of Fig.~\ref{fig:experimental_data}A).  This is consistent with the degeneracy structure of the GTDMs, which allows for a choice of basis functions that break the structure's five-fold rotational symmetry and localize on each of the five symmetry axes.  Next, we use a half-wave plate (HWP) to rotate the input polarization angle, with another linear polarizer at relative angle $\theta_2$ placed at the end face (see Materials and Methods).  As the HWP is rotated by 90 degrees (which causes the input polarization to rotate by 180 degrees), the output intensity varies sinusoidally over one cycle, and this curve shifts by a half-cycle if $\theta_2$ increases by 90 degrees (Fig.~\ref{fig:experimental_data}B).  We then fix the HWP at an extremal intensity setting (vertical dashes in Fig.~\ref{fig:experimental_data}B) and measure the intensity along a radial line passing through the output spot (cyan dashes in the inset of Fig.~\ref{fig:experimental_data}A), obtaining the profile shown in 
Fig.~\ref{fig:experimental_data}C.  We hence conclude that the polarization of the GTDM is preserved as it passes through the TPCF.  Moreover, upon tuning the wavelength filter, we find that this property is relatively insensitive to the operating wavelength (Fig.~\ref{fig:experimental_data}D).

A guided mode in a PCF should also be robust against coupling to cladding modes (because those can, in turn, couple to free space).  The GTDMs are advantageous in this respect, as they are tied to spectral charges originating from topological band invariants that do not change continuously when the system is weakly perturbed \cite{OzawaReview2019, TangReview2022}.  The bulk-defect correspondence demands that this charge be localized to the disclination center, thereby obstructing mode delocalization via the hybridization of GTDMs with bulk states.  This reasoning is also consistent with the nonzero local gap and index jump revealed by the spectral localizer (see the Supplementary Materials, Figs.~S3 and S4).  Although such ``topological protection'' is never absolute (e.g., it can be spoiled by finite-size effects), the TPCF evidently operates in a regime where the GTDMs indeed couple very weakly to other modes, as seen in the sizable gap between their dispersion relation and the cladding light line (Fig.~\ref{fig:band_indicators_charges}A) and their strong spatial localization (Fig.~\ref{fig:TPCF}E).

Accordingly, we expect the TPCF to perform well even when physically deformed.  To test this, we subject a $175$~$\textrm{m}$ long TPCF to two-loop bend, with $1$~$\textrm{cm}$ bending radius.  Over much of the operating wavelength range, we find that the transmittance is only slightly reduced relative to the straight TPCF (Fig.~\ref{fig:experimental_data}E).

For comparison, we fabricate a solid-core PCF with similar core size and an air hole radii/pitch ratio of $0.38$ (Fig.~\ref{fig:experimental_data}F, inset).  For each fiber type, we measure the difference in output power (in dBm) between the straight and bent fiber (again using a two-loop bend of radius $1$~$\textrm{cm}$).  The solid-core PCF is found to have much stronger bending losses, by up to 25~dBm, particularly in the $600$--$800$~$\mathrm{nm}$ range (Fig.~\ref{fig:experimental_data}F). To help understand these results, we perform numerical simulations in which the bend is modeled with a conformal transformation of the refractive index profile  (see the Supplementary Materials, section S2).  For the TPCF, the calculated $Q$ factors for the GTDMs remain almost the same with decreasing bending radius, down to a radius of around $2$~$\textrm{cm}$; by contrast, the solid-core PCF mode's $Q$ declines quickly as the bending radius goes below $4$~$\textrm{cm}$ (Fig.~\ref{fig:experimental_data}G).

\section*{Discussion}

We have realized a topological photonic crystal fiber (TPCF) supporting efficient broadband transmission that remains robust under strong bending.  The design is based on a topological crystalline insulator hosting localized disclination states due to a bulk-defect correspondence. Such topological states have never previously been implemented in optical fibers, despite having been studied in the context of photonic crystal slabs \cite{Peterson2021, Liu2021, Lin2023NRP}.  They turn out to have specific features that are advantageous for fiber waveguiding.  In particular, the fact that they can exist outside bandgaps, previously regarded as a relatively obscure quirk \cite{Peterson2021, Liu2021}, now enables the guided modes to reside below the cladding light line and thus decouple from the bulk modes.  We have presented experimental results indicating that the TPCF is more robust to bending losses than a comparable solid-core PCF based on a conventional design.  This calls for further quantitative studies, including comparisons to other fiber types as well as optimizations to our TPCF design, to determine whether topological modes are indeed advantageous for optical fibers.  It would also be interesting to explore alternative TPCF structures, including those employing different kinds of band topology; many of the frameworks we have used to analyse our TPCF, especially the spectral localizer, may be useful for guiding such work.

The built-in degeneracy of the topological modes in our TPCF offers particularly intriguing possibilities for further research. In the future, it should be possible to develop more optimized procedures for selectively addressing the ten degenerate topological modes, for the purposes of spatial division multiplexing \cite{Uden2014}, or studying their interaction with nonlinear effects in fibers \cite{SylvestreReview2021}.  It may also be possible to perform braiding on the degenerate topological modes, which would allow information to be encoded in their associated holonomy \cite{Chen2022braiding}.

While this manuscript was in submission, we noticed a work demonstrating topological states in a helically twisted fiber \cite{Beravat2016}. Unlike the disclination states utilized in this work, those topological states are edge states circulating around the boundary of the fiber core \cite{Roberts2024}.

\section*{Materials and Methods}

%%%%%%%%%%%%%%%% MATERIALS AND METHODS %%%%%%%%%%%%%%%

%\subsection*{Materials and Methods}

%The Materials and Methods section should contain details of the samples measured,
%experiments performed, observations taken, simulations run, data analysis, statistical methods etc.
%Give enough detail for any competent researcher in your field to fully reproduce the results.
%
%To refer to this section from the main text, use the numbered note in the reference list \cite{methods}.
%Refer to figures and tables in the same way as in the main text but now all capitalized e.g.
%Fig.~\ref{fig:example}, Table~\ref{tab:example},
%Fig.~\ref{fig:sup_example} and Table~\ref{tab:sup_example}.
%Cite references in the usual way \cite{example2},
%including any that are only cited in the supplement \cite{sm_example,conference_example}.
%
%The numbering of figures, tables, equations and pages has been reset to start from S1, as in
%\begin{equation}
%	\cos(2\theta) = \cos^2\theta - \sin^2\theta.
%	\label{eq:sup_example} % Use a logical label
%\end{equation}
%\subsubsection*{S1.~Design and fabrication details}
\subsection*{Experimental setup}
\label{secE}
The optical fibers are illuminated using a supercontinuum laser (YSL Photonics SC-PRO 7 Supercontinuum Source, wavelength range $430$--$2400\,\textrm{nm}$, peak bandwidth $1050$--$1080\,\textrm{nm}$).  The laser light passes through a lens and couples into the fiber head, which is fixed on a 3D stage.  To optimize the input coupling to the fiber core, the fiber tail is connected to a power meter and the 3D stage is moved along three dimensions until the measured signal reaches a maximum.  The power meter is then replaced by an optical signal analyser (Yokogawa AQ6370C, wavelength range $600$--$1700\,\textrm{nm}$) to record the output spectrum, or a camera beam profiler (Thorlabs BC106N-VIS/M, wavelength range $350$--$1100\,\textrm{nm}$) to record the mode profile.

To obtain the fiber loss spectrum (Fig.~1G, lower plot), we follow a standard cut-and-measure procedure.  The output spectrum is first measured for a TPCF of length 67~m. The fiber is then cut to a length of 15 m, and the output spectrum is measured again. The two spectra, expressed in logarithmic units, are subtracted from each other and the result divided by the truncated length to give the loss spectrum. The fiber transmittance (Fig.~1G, upper plot) is obtained by taking the ratio of the TPCF output power to the output power from a multimode fiber (MMF) launched by the same source. The two output powers are expressed in linear scale and the MMF output is self-normalized.  This transmittance is then converted to a logarithmic scale and plotted in units of dBm.

When characterising the polarization of the transmitted light (Figs.~3A--D), a tunable filter (Fianium LLTF Contrast SWIR, wavelength range $1000\,\textrm{nm}$--$2300\,\textrm{nm}$) is placed right after the source.  By adjusting the 3D stage on which the fiber head is mounted, we locate a setting in which the output intensity is concentrated on one of the five high-index regions (Fig.~3A, inset), and we then use this to obtain the results in Figs.~3B--D. The reference angle $\theta_2$ is arbitrarily chosen, but is fixed during all subsequent measurements. The intensities are directly extracted from the beam profiler.

In the bending loss experiment (Figs.~3E and F), we place a 90~m length fiber on a bending base with a preset bending radius.  The other experimental procedures are as previously stated.

\clearpage % Clear all remaining figures and tables then start a new page

%%%%%%%%%%%%%%%% ACKNOWLEDGEMENTS %%%%%%%%%%%%%%%

\section*{Acknowledgments}

We are grateful to Guoqing Chang for helpful discussions. 
\paragraph*{Funding:}
Y.~C.~acknowledges support from the Singapore National Research Foundation (NRF) under Competitive Research Program (CRP) NRF-CRP23-2019-0005 and the NRF Investigatorship NRF-NRFI08-2022-0001. Y.~C.~and Q.~J.~W.~acknowledge support from NRF-CRP23-2019-0007 and NRF-CRP29-2022-0003. Q.~J.~W.~acknowledges support from the Medium Sized Centre (MSC) Grant NRF-MSG-2023-0002 and the Singapore A*STAR Grant No.~A2090b0144. W.~C.~acknowledges support from the Singapore Ministry of Education (MOE) MOE-T2EP50122-0019. S.~W.~acknowledges support from the Laboratory Directed Research and Development program at Sandia National Laboratories. A.~C.~acknowledges support from the U.S.~Department of Energy, Office of Basic Energy Sciences, Division of Materials Sciences and Engineering. This work was performed in part at the Center for Integrated Nanotechnologies, an Office of Science User Facility operated for the U.S. Department of Energy (DOE) Office of Science. Sandia National Laboratories is a multimission laboratory managed and operated by National Technology \& Engineering Solutions of Sandia, LLC, a wholly owned subsidiary of Honeywell International, Inc., for the U.~S.~DOE's National Nuclear Security Administration under Contract No.~DE-NA-0003525. The views expressed in the article do not necessarily represent the views of the U.S.~DOE or the United States Government. 

\paragraph*{Author contributions:} 
B.Z. and Y.C. conceived the idea. B.Z. designed the fiber, performed the simulations and theoretical analysis. K.H. and W.C. performed the fiber fabrication. K.H. and B.Z. conducted the experiments. S.W. and A.C. performed the spectral localizer calculations. B.Z., S.W., A.C. and Y.C. wrote the manuscript. Y.W., R.B., H.X. and Q.W. discussed and commented on the manuscript. Q.J.W., W.C. and Y.C. supervised the project.

\paragraph*{Competing interests:}
There are no competing interests to declare.

\paragraph*{Data and materials availability:}
All data needed to evaluate the conclusions in the paper are present in the paper and/or the Supplementary Materials.

%%%%%%%%%%%%%%%% SUPPLEMENT LIST %%%%%%%%%%%%%%%

% List the contents of your Supplementary Materials, including the numbers of any
% supplementary figures, tables, external data files etc. and any references that are
% cited only in the supplement. In this example, refs. 7-8 are cited only in the supplement.
% Fill out your numbers accordingly and delete any lines that aren't applicable.
\subsection*{Supplementary materials}
Supplementary Text S1 to S5\\
Figs. S1 to S4\\
References \textit{(56-\arabic{enumiv})}\\ % automatically fills out the last reference number
% (filling out the other numbers automatically is possible but fiddly and liable to break)

%%%%%%%%%%%%%%%% FIGURES AND CAPTIONS %%%%%%%%%%%%%%%

\section*{Figures and Tables}

\begin{figure}
  \centering
  \includegraphics[width=1\textwidth]{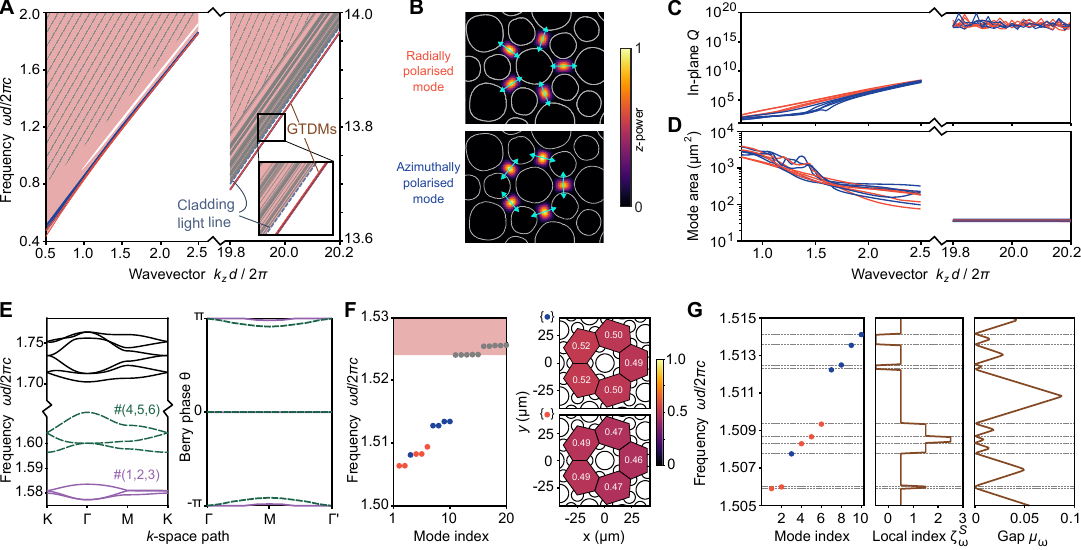}
  \caption{\textbf{Analysis of guided topological defect modes (GTDMs).} (\textbf{A}) Band diagram for the measured TPCF structure (Fig.~\ref{fig:TPCF}D).  The GTDMs are plotted in red (radially polarized) and blue (azimuthally polarized), and bulk band frequencies are drawn as pink areas.  In the large-$k_z$ regime on the right, dispersion curves for individual bulk states are plotted in gray.  The gray-stroked region contains numerous modes that cannot be resolved numerically. (\textbf{B}) Calculated intensity profiles (normalized power flow in $z$) for exemplary radially and azimuthally polarized GTDMs at $k_z d/2\pi = 20$. Polarization directions are indicated by cyan arrows. (\textbf{C}) In-plane quality ($Q$) factors of the GTDMs versus $k_z$. (\textbf{D}) Mode areas of the GTDMs versus $k_z$. (\textbf{E}) Characterization of bulk TCI bands at fixed $k_z$.  Left panel: band spectrum for the periodic TCI structure (corresponding to Fig.~\ref{fig:TPCF}B, upper plot).  Right panel: Berry phases of the Wilson loop operator for different base points, using bulk bands $\#$(1, 2, 3) and $\#$(4, 5, 6).  (\textbf{F}) Calculated eigenfrequencies for the preform hole profile (left panel; pink areas denote bulk bands), and the corresponding spectral charges in the five unit cells around the center (right panel).  Blue/red dots respectively indicate azimuthally/radially polarized GTDMs. For details, see the Supplementary Materials, section S3.  (\textbf{G}) Topological characterization via the spectral localizer (see the Supplementary Materials, section S4).  Left panel: eigenfrequencies of a symmetrized structure based on the preform profile. Center panel: the local index $\zeta_\omega^{S}$, where $S$ is mirror symmetry around the $x$ axis.  Right panel: the local gap measure $\mu_\omega$. The results for E--G are calculated at $k_z d/2\pi=2$.}
  \label{fig:band_indicators_charges}
\end{figure}

\begin{figure}
  \centering
  \includegraphics[width=1\textwidth]{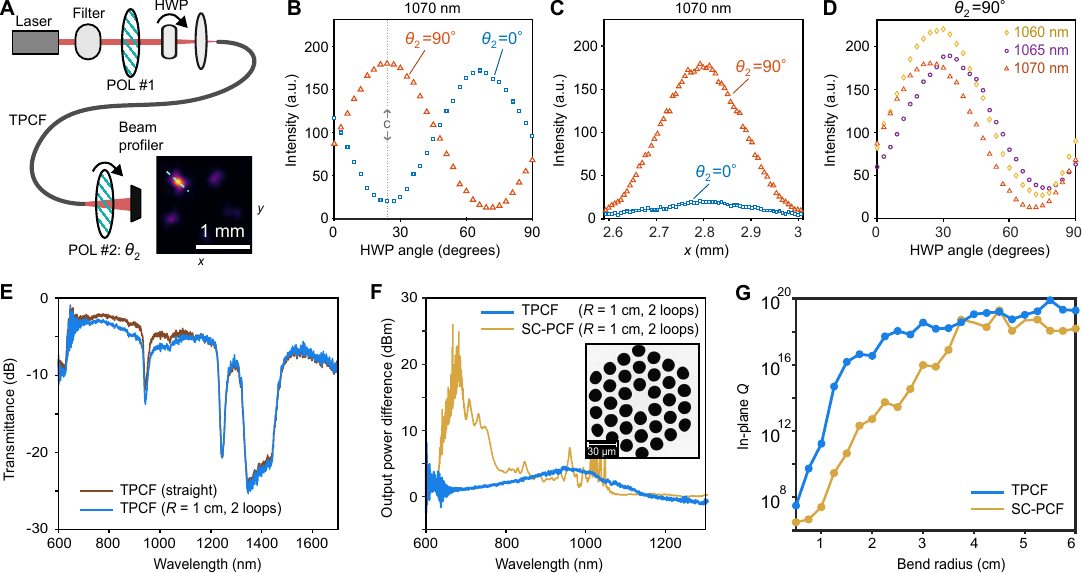}
  \caption{\textbf{Polarization dependence and bending resistance.} (\textbf{A}) Experimental setup with a tunable filter, fixed linear polarizer, and rotatable half-wave plate (HWP) between the source and the input face of a 0.5~m TPCF.  A second linear polarizer, with variable angle $\theta_2$, is placed between the end face and a beam profiler.  Inset: a measured output intensity profile concentrated at one of the high-index regions around the central air hole.  (\textbf{B}) Measured intensity at the center of the selected spot, versus HWP angle.  Results are shown for two values of $\theta_2$ differing by $90^\circ$, with fixed wavelength 1070~nm.  (\textbf{C}) Variation of intensity with position, measured along a radial line passing through the selected spot (cyan dashes in the inset of (A)), with HWP fixed at $24^\circ$ (vertical dashes in (B)). (\textbf{D}) Intensity at the center of the selected spot versus HWP angle, for three different input wavelengths and fixed $\theta_2 = 90^\circ$. (\textbf{E}) Measured transmittance for a straight TPCF (brown) and a TPCF with a two-loop bend of radius 1~cm (blue). (\textbf{F}) Output power difference between a straight fiber and one with a two-loop bend of radius 1~cm, for the TPCF (blue) and a comparable solid-core (SC) PCF (yellow).  Inset: Scanning electron microscope image of the SC-PCF. (\textbf{G}) Calculated mean in-plane $Q$ factors for the GTDMs in the TPCF (blue) and the fundamental core modes in the SC-PCF (yellow), for different bending radii at $k_z d/2\pi=30$ (corresponding to $\sim$740~nm).  For the SC-PCF in (F) and (G), the ratio between the air hole radii and the pitch is $0.38$.}
  \label{fig:experimental_data}
\end{figure}

%%%%%%%%%%%%%%%% END OF MAIN TEXT %%%%%%%%%%%%%%%

\newpage

%%%%%%%%%%%%%%%% START OF SUPPLEMENT %%%%%%%%%%%%%%%

% Figures, tables, equations and pages in the supplement are numbered S1, S2 etc.
\renewcommand{\thefigure}{S\arabic{figure}}
\renewcommand{\thetable}{S\arabic{table}}
\renewcommand{\theequation}{S\arabic{equation}}
\renewcommand{\thepage}{S\arabic{page}}
\setcounter{figure}{0}
\setcounter{table}{0}
\setcounter{equation}{0}
\setcounter{page}{1} % not 0 as \newpage already started a supplementary page
% References continue the numbering from the main text.

%%%%%%%%%%%%%%%% SUPPLEMENT TITLE PAGE %%%%%%%%%%%%%%%

\begin{center}
\section*{Supplementary Materials for\\ \scititle}

% Author list for the supplement
% Indicate the corresponding authors, but do NOT include institutions here
% It would be nice if the template auto-generated this, but doing so is complicated...
Bofeng Zhu$^{\dagger}$,
Kevin Hean$^{\dagger}$,
Stephan Wong,
Yuxi Wang,
Rimi Banerjee,
Haoran Xue,
Qiang Wang,
Alexander Cerjan,
Qi Jie Wang$^{\ast}$,
Wonkeun Chang$^{\ast}$,
Yidong Chong$^{\ast}$,\\
\small$^\ast$Corresponding author. Email: qjwang@ntu.edu.sg, wonkeun.chang@ntu.edu.sg, yidong@ntu.edu.sg\\
% Joint contributions can be indicated like this
\small$^\dagger$These authors contributed equally to this work.

\end{center}

% Fill out the numbers for each type of supplementary material,
% and delete any lines that aren't applicable.
% These are just example numbers that don't match the rest of this template.
\subsubsection*{This PDF file includes:}
Supplementary Text S1 to S5\\
Figs.~S1 to S4\\

\newpage

%%%%%%%%%%%%%%%% MATERIALS AND METHODS %%%%%%%%%%%%%%%

%\subsection*{Materials and Methods}
\subsection*{Supplementary Text}
%The Materials and Methods section should contain details of the samples measured,
%experiments performed, observations taken, simulations run, data analysis, statistical methods etc.
%Give enough detail for any competent researcher in your field to fully reproduce the results.
%
%To refer to this section from the main text, use the numbered note in the reference list \cite{methods}.
%Refer to figures and tables in the same way as in the main text but now all capitalized e.g.
%Fig.~\ref{fig:example}, Table~\ref{tab:example},
%Fig.~\ref{fig:sup_example} and Table~\ref{tab:sup_example}.
%Cite references in the usual way \cite{example2},
%including any that are only cited in the supplement \cite{sm_example,conference_example}.
%
%The numbering of figures, tables, equations and pages has been reset to start from S1, as in
%\begin{equation}
%	\cos(2\theta) = \cos^2\theta - \sin^2\theta.
%	\label{eq:sup_example} % Use a logical label
%\end{equation}

\subsubsection*{S1.~Design and fabrication details}
\label{secF}

\begin{figure}
  \centering
  \includegraphics[width=1\textwidth]{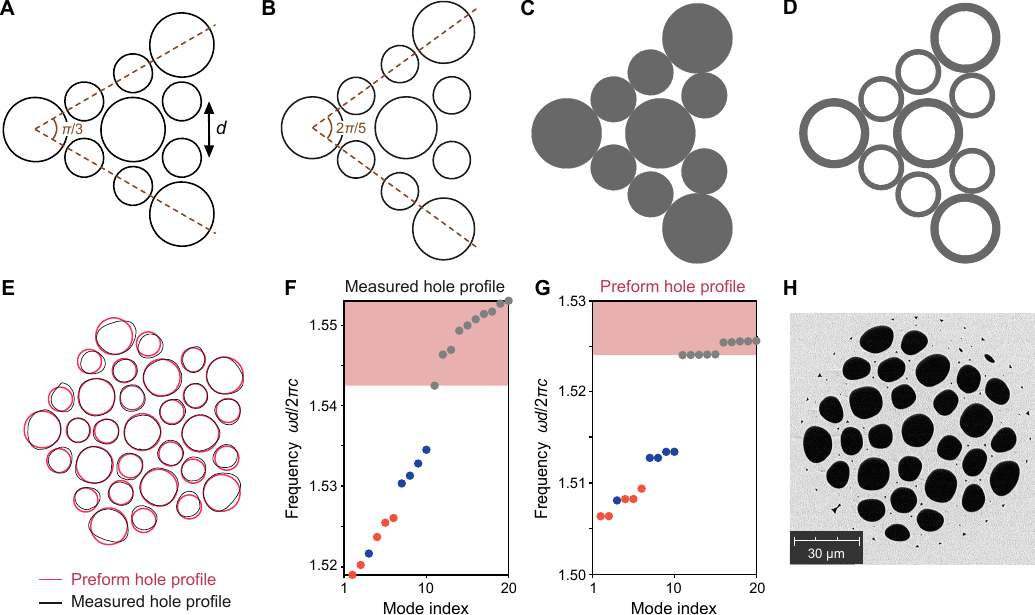}
  \caption{\textbf{Lattice design.} (\textbf{A}) Wedge with opening angle $\pi/3$, extracted from a triangular lattice with nearest neighbor center-to-center spacing $d$.  The discs (air holes) have alternating radii of $0.57d$ and $0.35d$.  (\textbf{B}) Wedge with opening angle $2\pi/5$, generated from (A) by uniformly scaling the azimuthal coordinates of the site centers.  This corresponds to the structure shown in Fig.~1A.  (\textbf{C}) Wedge generated from (B) by gradually enlarging the disc radii and adjusting their positions, with boundaries fixed, until they are jam-packed. (\textbf{D}) Structure generated from (C) by assigning inner air holes of radius $R_1' = 0.49d$ and $R_2' = 0.33d$ to the discs.  This corresponds to the target stacking arrangement shown in Fig.~1C. (\textbf{E}) Comparison between the preform hole profile based on (D) (black lines), and the measured hole profile of the fabricated TPCF (pink lines). The air holes radii are slightly enlarged to $R_1 = 0.55d$ and $R_2 = 0.36d$. (\textbf{F} and \textbf{G}) Calculated eigenfrequencies at $k_z d/2\pi = 2$ for the measured hole profile (F) and the preform hole profile (G).  Red (blue) data points correspond to GTDMs with radial (azimuthal) polarization, while gray data points are bulk states.  The pink-shaded region indicates the bulk band. (\textbf{H}) Scanning electron microscope image of the fiber cross section.  The small black specks correspond to non-filled interstitial air holes, which can be seen to mostly occur in the out-of-core region.}
  \label{fig:extended_data_0}
\end{figure}

To design the TPCF structure, we begin with a wedge of opening angle $\pi/3$  extracted from the periodic TCI structure (Fig.~\ref{fig:extended_data_0}A).  This has circular air holes of alternating radii $0.57d$ and $0.35d$, where $d$ is the center-to-center distance between neighbouring holes.  We deform this into a $2\pi/5$ wedge by scaling each site's azimuthal coordinate by $6/5$ (Fig.~\ref{fig:extended_data_0}B).  Copying this wedge to the remaining lattice sectors yields the structure of Fig.~1A, featuring a disclination of Frank angle $-\pi/3$.

Next we make small adjustments to the site positions, with the aim of improving the overall spatial uniformity of the PCF structure.  This is achieved with the aid of a molecular dynamics (\textsc{LAMMPS} \cite{lammps}) simulation, which moves a set of ``atoms'' centered at the site positions while also gradually enlarging their radii, subject to the fixed boundaries of the wedge \cite{Yang2011jamming}.  This yields a jam-packed configuration (Fig.~\ref{fig:extended_data_0}C), in which the atomic radii ($R_1'' = 0.65d$ and $R_2'' = 0.43d$) correspond to the outer surfaces of the glass capillaries we aim to use in the fiber preform (see below).  Finally, we define the air holes by downscaling the radii to $R_1' = 0.49d$ and $R_2' = 0.33d$, corresponding to the inner surfaces of the capillaries.  The resulting structure (Fig.~\ref{fig:extended_data_0}D) corresponds to the arrangement shown in the inset of Fig.~1C.

To fabricate the fiber, we draw glass tubes of outer (inner) diameter 25~mm (19~mm) into two sets of smaller tubes, or capillaries.  After adjusting the drawing conditions (temperature, vacuum pressure, etc.), we obtain capillaries with outer diameters $R_1' = 3.33\,\textrm{mm}$ (11 pieces) and $R_2' = 2.18\,\textrm{mm}$ (20 pieces), close to the ideal ratio described above.  Additional glass tubes of outer (inner) diameter 10~mm (3~mm), composed of the same silica material as the capillaries, are drawn into solid rods with collapsed inner air holes and outer diameters of 1.095~mm (5 pieces), 0.702 mm (20 pieces), 1.619~mm (15 pieces), and 1.905~mm (10 pieces).  The capillaries and rods are stacked in a jacket of outer (inner) diameter 25~mm (19~mm) according to the arrangment in the inset of Fig.~1C, with the rods filling the major gaps between the capillaries.  The preform is drawn into a preform cane of diameter 4.7~mm (Fig.~1D), which is in turn drawn into the TPCF in a fiber drawing tower.

The drawing process introduces deformations to the dielectric structure.  From a scanning electron microscope image of the fiber end face (Fig.~1D), we find that the holes become non-circular, but their positions are relatively unaffected (Fig.~\ref{fig:extended_data_0}E).  The air hole radii are also slightly enlarged to $R_1 = 0.55d$ and $R_2 = 0.36d$, which are the values used in the preform hole profile for numerical simulations (see below). From numerical calculations of the eigenmodes (see below), we find that the deformations induce small shifts in the frequencies of the GTDMs and bulk states, but do not alter the qualitative features of the spectrum (Fig.~\ref{fig:extended_data_0}F and G). We also observe that some of the interstitial air holes fail to be completely filled-in, resulting in some small holes that are mainly concentrated in the out-of-core areas (Fig.~\ref{fig:extended_data_0}H); we expect that these imperfections can be eliminated in the future by optimizing the drawing conditions, but they are in any case unlikely to influence the GTDMs, which are localized near the core.

We use the TPCF's measured hole profile (i.e., the black lines in Fig.~\ref{fig:extended_data_0}E) for most of our numerical calculations, including the band diagram (Fig.~2A), various GTDM properties (Fig.~2B--D), bending performance calculation (Fig.~3G) and 2D spectral localizer calculations (Fig.~\ref{fig:extended_data_3}). The exception is the spectral charge calculation (Fig.~2F), which uses the preform hole profile, and the 1D spectral localizer calculations (Fig.~2G and Fig.~\ref{fig:extended_data_2}B), which use the preform hole profile with a mirror symmetry-breaking perturbation (see below).

The conventional solid-core PCF is fabricated via the same stack-and-draw process, using a single set of glass tubes with outer (inner) radius 20~mm (16~mm) drawn into capillaries of outer diameter 2.28~mm (36 pieces).  A single solid rod of 2.28~mm is used for the core.  The hexagonal preform is supported in the cylindrical jacket by solid rods of outer diameter 0.61~mm (6 pieces), 0.75~mm (30 pieces) and 0.97~mm (12 pieces).

\subsubsection*{S2.~Calculating photonic eigenmodes and band diagrams}
\label{secG}
An optical fiber is translationally symmetric in the axial ($z$) direction.  For a given axial wavenumber ($k_z$), the electromagnetic eigenmodes can be determined by solving Maxwell's equations in the 2D ($x$-$y$) plane.  In this work, we perform the calculations numerically via the finite-element method (FEM) simulation software COMSOL Multiphysics, solving the full vectorial form of Maxwell's equations with no additional approximations apart from the discretization of space.  The principal input to this calculation is the real-valued dielectric profile $\epsilon(\mathbf{r})$, where $\mathbf{r} = (x,y)$ is the position in the 2D plane.  We model the structure's high-index (silica) and low-index (air) regions with $\epsilon=2.1$ and $\epsilon=1$ respectively.

Each calculated eigenmode $\mu$ has some angular frequency $\omega_\mu$ and electric field profile
\begin{equation}
  \mathbf{E}_\mu(\mathbf{r}) \equiv \langle \mathbf{r} | \mu \rangle,
\end{equation}
which is a complex 3-vector-valued field defined in the 2D space; from this, the full physical electric field is given by $\mathrm{Re}\left[\mathbf{E}_\mu(\mathbf{r}) \exp(ik_z z)\right]$.  Both $\omega_\mu$ and $\mathbf{E}_\mu(\mathbf{r})$ depend implicitly on $k_z$. The inner product between two normalized eigenmodes is defined as \cite{Joannopoulos2008}
\begin{equation}
  \langle \mu | \nu\rangle = \int d^2 r
  \; \epsilon(\mathrm{r}) \; \mathbf{E}_\mu^*(\mathbf{r}) \cdot
  \mathbf{E}_{\nu}(\mathbf{r}).
  \label{innerprod}
\end{equation}
If the boundary conditions are Hermitian (e.g., Dirichlet or periodic boundary conditions), Maxwell's equations consitute a Hermitian generalized eigenproblem, so the orthogonality relation $\langle \mu | \nu\rangle = \delta_{\mu\nu}$ holds \cite{Joannopoulos2008}.

When simulating the spatially finite TPCF structure, our choice of boundary conditions depends on the circumstances.  For the spectral localizer (Fig.~2G and Fig.~\ref{fig:extended_data_2}B) and finding the spectral charges (Fig.~2F), it is important that the eigenproblem be Hermitian, so we apply perfect electric conductor (PEC) boundary counditions to the system's exterior boundary.  In all other cases, including calculations of the mode profiles (Fig.~1E and Fig.~2B), dispersion relations (Fig.~2A) and the 2D localizer calculations (Fig.~\ref{fig:extended_data_3}), we apply impedance boundary conditions, which are equivalent to an infinite external medium of $\epsilon=2.1$.  Since light can leak out of the fiber and escape to infinity, the eigenfrequencies become complex; the real parts are used in the dispersion relations (Fig.~2A), and the imaginary parts are used to determine $Q$ factors (Fig.~2C and Fig.~3G).  From the field distributions, we can also extract the effective mode areas (Fig.~2D), which are defined as \cite{MarkosReview2017}
\begin{equation}
  A = \frac{\displaystyle \left(\int d^2 r |E_{\mu}(\textbf{r})|^2 \right)^2}{\displaystyle\int d^2 r |E_{\mu}(\textbf{r})|^4},
\end{equation}
where the eigenmode $E_\mu(\mathbf{r})$ has been normalized using the inner product Eq.~\eqref{innerprod}.

When modeling the effects of bending the fiber (Fig.~3G and Fig.~\ref{fig:extended_data_3}), we perform a conformal transformation on the refractive index profile \cite{Poletti2014}.  The modified refractive index (the square root of the dielectric constant) is
\begin{equation}
  n'(x,y) = n(x,y) \, \left(1-\frac{x}{R}\right),
\label{eq:ri_bending}
\end{equation} 
where $n(x,y)$ is the refractive index distribution for the straight fiber with $R$ as the bending radius. The ideal preform hole profile (Fig.~\ref{fig:extended_data_0}E) with such index modulation is used for the spectral localizer calculations presented in Fig.~2G and Fig.~\ref{fig:extended_data_2}B.  We use the measured hole profile, which lacks a mirror symmetry, for the additional spectral localizer analysis based on local gap profiles in Fig.~\ref{fig:extended_data_3}.

When analysing the underlying photonic crystal structure (Fig.~1B), which is spatially infinite and periodic in the $x$-$y$ plane, we impose Bloch's theorem by writing
\begin{equation}
  \mathbf{E}_{n,\mathbf{k}}(\mathbf{r})
  = \mathbf{u}_{n,\mathbf{k}}
  (\mathbf{r}) \; e^{i\mathbf{k}\cdot\mathbf{r}},
\end{equation}
where $\mathbf{k}$ is a 2D wave-vector (quasimomentum) and $n$ is a band index.  Such eigenfunctions are calculated numerically using a single unit cell with Floquet-Bloch boundary conditions.  The eigenproblem is Hermitian, with inner products between Bloch states given by
\begin{equation}
  \langle n,\mathbf{k} | n',\mathbf{k}' \rangle = \int_{\mathrm{u.c.}} d^2 r
  \; \epsilon(\mathrm{r}) \; \mathbf{u}_{n,\mathbf{k}}^*(\mathbf{r}) \cdot
  \mathbf{u}_{n',\mathbf{k}'}(\mathbf{r}),
  \label{Bloch_product}
\end{equation}
where the integral is taken over a single unit cell.

%\textbf{f}, Phase profiles of the out-of-plane electric field ($E_z$) for the six Bloch states marked in \textbf{e} at high-symmetry momentum points. 

%%%%%%%%%%%%%%%% SUPPLEMENTARY TEXT %%%%%%%%%%%%%%%

%\subsection*{Supplementary Text}
%The Supplementary Text section can only be used to directly support statements made in the main text
%e.g. to present more detailed justifications of assumptions, investigate alternative scenarios,
%provide extended acknowledgements etc.
%Material in this section cannot claim results or conclusions that weren't mentioned in the main text.
%To refer to this section from the main text, just write (Supplementary Text).

\subsubsection*{S3.~Wannier centers and spectral charge}
\label{secH}

\begin{figure}
  \centering
  \includegraphics[width=1\textwidth]{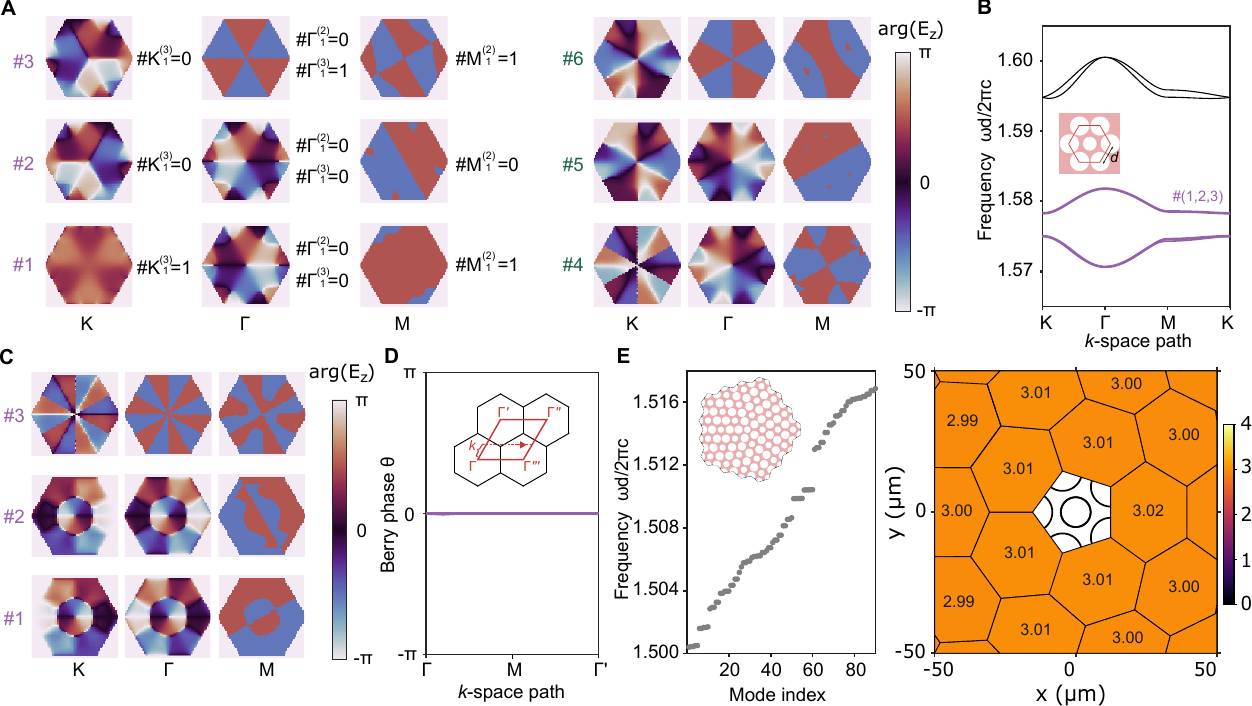}
  \caption{\textbf{Wilson loop and spectral charge calculations.} (\textbf{A}) Phase profiles of the out-of-plane electric field ($E_z$) for the six Bloch states marked in Fig.~2E at high-symmetry momentum points. The symmetry indicators of bulk bands $\#(1,2,3)$ are labeled on the right. (\textbf{B}) Lowest five bands for the periodic photonic structure with trivial crystalline topology, calculated from the unit cell in Fig.~1B (lower plot). (\textbf{C}) Phase profiles of the out-of-plane electric field ($E_z$) for the Bloch states $\#(1,2,3)$ in the trivial structure of (\textbf{B}) at high-symmetry momentum points. (\textbf{D}) Berry phases of the Wilson loop operator at different initial $k$ for the trivial bulk bands $\#(1,2,3)$. Inset: the schematic of the Wilson loop. The base point lies along the path $\Gamma \rightarrow \Gamma'$, and the Wilson loop follows the dashed line running parallel to $\Gamma$--$\Gamma'''$. (\textbf{E}) Left panel: calculated eigenfrequencies for the trivial PCF structure, based on Fig.~2F but with large and small air holes switched. Inset: schematic of the lattice, composed of 30 intact unit cells and a central defect. Right panel: calculated spectral charges (black numbers) for the schematic structure in the left panel. In (\textbf{A--E}), the eigenmodes are calculated at $k_z d/2\pi = 2$. The spectral charge calculations are performed using symmetric preform hole profiles.}
  \label{fig:extended_data_1}
\end{figure}

Like other TCIs, the $C_{6}$-symmetric precursor structure for our TPCF can be topologically characterized using symmetry indicators.  In this section, we will explain the procedure, which largely follows earlier studies \cite{Benalcazar2019, Li2020, Liu2021}.  It is worth noting that these methods are applicable even though the photonic structure we are considering is not based on a tight-binding model.

Given a set of bands that are well-defined (i.e., not overlapping with any other bands), one defines $\#\Pi_{p}^{(P)}$ as the number of bands at a high-symmetry momentum point $\Pi \in \{\mathrm{\Gamma},\mathrm{M},\mathrm{K}\}$ that has the $C_n$ rotation eigenvalue $e^{\mathrm{i}2\pi(p-1)/P}$ ($p=1,\dots,P$).  We can then calculate the topological index $(\chi_\mathrm{M},\chi_\mathrm{K})$, where $\chi_\mathrm{M}= \#\mathrm{M}_{1}^{(2)} - \#\Gamma_{1}^{(2)}$ and $\chi_\mathrm{K}= \#\mathrm{K}_1^{(3)} - \#\Gamma_1^{(3)}$ \cite{Benalcazar2019}.

From their Bloch function phase profiles, the topological index of bands $\#(1,2,3)$ is calculated by first calculating the $\Pi_{p}^{(n)}$ index for each individual band and then summing these indices across all three bands (see Fig.~\ref{fig:extended_data_1}A),

 \begin{align}
   \begin{aligned}
   \#\Gamma_1^{(2)} &= 0\,(0), \\ 
   \#\mathrm{M}_1^{(2)} &= 2\,(2)
   \end{aligned}\quad
   \begin{aligned}
     \#\Gamma_1^{(3)} &= 1\,(1) \\    
     \#\mathrm{K}_1^{(3)} &= 1\,(1).
   \end{aligned}
 \end{align}
The index of bands $\#(4,5,6)$ can be calculated in a similar way, and both sets of bands have index $(\chi_M,\chi_K)=(2,0)$. The indicators of topological trivial case can be calculated from Fig.~\ref{fig:extended_data_1}B and C, with a trivial index of $(\chi_M,\chi_K)=(0,0)$.

The topological bulk-defect correspondence in such TCIs states that the symmetry indicators are related to defect charges of \cite{Li2020,Liu2021}
 \begin{equation} \label{eq:2}
   \mathrm{Q}_{\mathrm{dis}} = \frac{\Omega}{2\pi}\left(\frac{3}{2}\chi_{\mathrm{M}}-\chi_{\mathrm{K}}\right)
   \,\,\mathrm{mod\,\,1},
\end{equation}
where $\Omega=-\pi/3$ is the Frank angle of the disclination.  Applying this to our TCI, the two sets of bands, $\#(1,2,3)$ and $\#(4,5,6)$, are each predicted to produce fractional charge $\mathrm{Q}_{\mathrm{dis}}=1/2~(\textrm{mod}~1)$.

To verify the correspondence, we explicitly calculate the spectral charges.  Even though there are two half-charges, we can distinguish them by exploiting the polarization structure of the GTDMs.  For each polarization (radial or azimuthal), we pick out the five GTDMs and calculate
\begin{equation} \label{charge}
  \sum_{u\,\in\,\mathrm{GTDMs}} \int_{S} d^2 r
  \; \epsilon(\mathrm{r}) \; \left | \mathbf{E}_\mu^*(\mathbf{r}) \right|^2,
\end{equation}
where $S$ is one of the five unit cells surrounding the core unit cell.  %For each polarization, we indeed find a charge of $\approx 0.5$ in each of these five areas (Fig.~\ref{fig:extended_data_1}\textbf{c}). 

To determine the Wannier centers of the photonic TCI (Fig.~1B), we use the Wilson loop approach \cite{Wang2019, Blanco2020}.  Given a set of degenerate bands and a path $k+l \leftarrow k$ passing across the Brillouin zone, we discretise the path into steps separated by $\Delta\mathbf{k}$ and calculate
\begin{align}
  W_{k+l \leftarrow k} &= G^{k+l-\Delta k} \, G^{k+l-2\Delta k} \cdots G^{k+\Delta k} \, G^k \label{eq:3} \\
  G^{k}_{mn} &= \langle\, m,\mathbf{k} \, |\,
  n, (\mathbf{k}+\Delta\mathbf{k})\,\rangle,
\label{eq:4}
\end{align}
where $m, n$ are band indices and Eq.~\eqref{eq:4} uses the inner product defined in Eq.~\eqref{Bloch_product}. To construct the path of the Wilson loop, we first consider a rhombus with corners at four adjacent $\Gamma$ points in the extended Brillouin zone (inset of Fig.~\ref{fig:extended_data_1}D).  The initial $k$-point (or ``base point'') is swept along the path $\Gamma \rightarrow \mathrm{M} \rightarrow \Gamma'$, with the Wilson loop path taken parallel to the line $\Gamma \rightarrow \Gamma'''$.   

The eigenvalue spectrum of the $W$ operator is adiabatically deformable to a set of centers of localized Wannier functions in real space.  As established in previous works \cite{Neupert2018, Vanderbilt2018, Blanco2020}, topologically nontrivial TCIs have nonzero values of the Berry phase $\mathrm{Im}\,\{\mathrm{log}\,\left[\mathrm{det}\left(W_{k+l \leftarrow k}\right)\right]\}$ (Fig.~2E), corresponding to the Wannier centers not being located at the center of the unit cell. In contrast, the Berry phase of trivial TCI remains close to zero, as shown in Fig.~\ref{fig:extended_data_1}D.

%As the initial $k$-point is varied, we observe nontrivial Berry phases remaining around $\pm \pi$ (Fig.~\ref{fig:extended_data_1}\textbf{b}), corresponding to having Wannier centers localized to the Wyckoff position 3\textbf{c} \cite{Blanco2020}.

The calculation of spectral charge for the topological trivial case (Fig.~\ref{fig:extended_data_1}E) is similar but the summation is done for all the 90 bulk states below the bandgap, which are contributed from the lowest three bulk bands, i.e., band \#(1,2,3) in Fig.~\ref{fig:extended_data_1}B, of all the 30 intact unit cells.

\subsubsection*{S4.~Symmetry-reduced spectral localizer}
\label{secI}

The calculations in the previous section strongly indicate that the TPCF should host topological disclination states, but there are some limitations in the argument.  First, those characterization frameworks refer to an underlying periodic bulk TCI, yet the symmetries on which the TCI relies are broken by the disclination and the various aforementioned deformations in the TPCF.  Second, in a lattice exhibiting charge fractionalization, it is sometimes possible for a strong lattice disturbance, like the core of a disclination, to remove the topological states expected to accompany the fractionalization \cite{Peterson2021}.  Finally, even if the lattice hosts topological disclination states, a given disclination state can still be accidental, i.e., non-topological.  To resolve these doubts, we turn to the spectral localizer framework \cite{Loring2015,Cerjan2022,JMP2023, Cerjan2024}, which provides a way to directly establish that the disclination states in the TPCF are topological in origin as well as robust.

For a system with a single relevant position operator $Y_c$, the spectral localizer characterizes the local topology at frequency $\omega$ and position $y$ by combining $Y_c$ and the system's Hamiltonian $H_{\text{eff},c}(\omega)$ into
\begin{equation}
\label{eq:localizer}
L_{(y,\omega)}(Y_c, H_\text{eff,c}) = 
\left(
\begin{array}{cc}
0 & H_\text{eff,c}(\omega) - i \kappa \left( Y_c - y \mathbf{1}_c \right)  \\
H_\text{eff,c}(\omega) + i \kappa \left( Y_c - y \mathbf{1}_c \right) & 0 
\end{array}
\right),
\end{equation} 
where $\kappa$ is a scaling coefficient used to make the units consistent and to balance the spectral weight between $Y_c$ and $H_\text{eff,c}$.
The notation here follows Ref.~\cite{Wong2024}.  Both $Y_c$ and $H_{\text{eff},c}$ are numerically extracted from the discretized master equations in the FEM eigenfrequency solver---the same solver used all in our other numerical calculations.  In particular, $H_{\text{eff},c}$ is a matrix-valued function of the frequency $\omega$, meaning the photonic eigenmodes correspond to eigenvectors of $H_\text{eff,c}(\omega)$ with zero eigenvalue \cite{Wong2024}.

The TPCF structure possesses a mirror symmetry  $M_y: y \rightarrow -y$, which satisfies $M_y^2 = \mathrm{\textbf{1}}$, $H_{\text{eff},c} M_y = M_y H_{\text{eff},c}$ and $Y_c M_y = - M_y Y_c$.  Using this symmetry, the reduced spectral localizer can be constructed as
\begin{equation}
\label{eq:reduced_localizer}
L_\omega^{M_y}(Y_c, H_\text{eff,c}) = \big[ H_\text{eff,c}(\omega) + i \kappa Y_c \big] M_y.
\end{equation} 
Topological defect modes can then be characterized by a local marker defined as
\begin{equation}
\label{eq:local_index}
\zeta_\omega^{M_y}(Y_c, H_\text{eff,c}) = \frac{1}{2} \text{sig} \left[ L_\omega^{M_y}(Y_c, H_\text{eff,c}) \right],
\end{equation} 
where $\text{sig}[\cdots]$ denotes the signature (i.e., the difference between the total number of positive and negative eigenvalues).  Note that Eq.~\eqref{eq:reduced_localizer} is Hermitian, so its signature is well-defined.

The key idea is to use the local index \eqref{eq:local_index} to classify the topology of the system, in terms of what kind of atomic limit it is continuable to, at each frequency $\omega$ and subject to the specified symmetry $M_y$.  For example, in a periodic lattice, one atomic limit would correspond to Wannier centers located at the center of the unit cell, while an obstructed atomic limit would correspond to Wannier centers located at the edges or corners of the unit cell.
%% Thus, distinct values of $\zeta_\omega^{M_y}$ correspond to distinct topological crystalline phases that cannot be continuously deformed into one other, unless the symmetry is broken or the local gap closes.

If the value of $\zeta_\omega^{M_y}$ changes at a frequency $\omega$, that implies that the modes at that frequency have non-trivial topology \cite{JMP2023}.  Such a change happens when an eigenvalue of the spectral localizer crosses zero.  Therefore, the local marker is also associated with a local measure of topological robustness via a ``local gap'',
\begin{equation}
\label{eq:local_gap}
\mu_\omega(Y_c, H_\text{eff,c}) = \text{min} \left| \text{spec} \left[ L_\omega^{M_y}(Y_c, H_\text{eff,c}) \right] \right|,
\end{equation}
where $\text{spec}[\cdots]$ denotes the spectrum of the matrix.

We apply the spectral localizer directly on the finite element represention of the TPCF's photonic structure (see Section~S3).  When doing this, there are two important subtleties to handle.  The first involves the boundary conditions encoded into the FEM matrices.  The subscript $c$ in $H_{\text{eff},c}$, $Y_c$ and $\mathbf{1}_c$ refers to ``eliminated matrices'', which have undergone a procedure whereby all degrees of freedom involved in the boundary conditions are removed \cite{Wong2024}.  This projection is realized using matrices denoted by ``$\text{Nullf}$'' and ``$\text{Null}$'', composed of basis vectors spanning the null space of the constraint force Jacobian matrix and the constraint Jacobian matrix, respectively: specifically, $H_{\text{eff},c} = \text{Nullf}^T H_\text{eff} \text{Null}$, $Y_c = \text{Nullf}^T Y \text{Null}$ and $\mathbf{1}_c = \text{Nullf}^T \text{Null}$, where $H_\text{eff}$ and $Y$ are matrices retrieved directly from the FEM (including boundary degrees of freedom).

The next subtlety involves the mirror symmetry.  Even if the structure and the mesh are both symmetric with respect to $M_y$, the discretized FEM master equations may not yield a (non-eliminated) effective Hamiltonian $H_\text{eff}$ commuting with $M_y$.  This can be problematic since, as explained above, the local marker assumes mirror symmetry, which is necessary to guarantee the Hermiticity of Eq.~\eqref{eq:reduced_localizer}.  To bypass this difficulty, we decompose $H_\text{eff}$ into even and odd subspaces with respect to $M_y$,
\begin{equation}
H_\text{eff} = 
\left(
\begin{array}{cc}
H_{\text{eff},+} & 0  \\
0 & H_{\text{eff},-} 
\end{array}
\right).
\end{equation}
The sub-Hamiltonians $H_{\text{eff},+}$ and $H_{\text{eff},-}$ are obtained from the FEM solver by applying symmetric/antisymmetric boundary conditions to the mirror line.  In a similar way, we decompose the $\text{Nullf}$ and $\text{Null}$ matrices.  In this basis, the mirror symmetry matrix reads
\begin{equation}
M_y = 
\left(
\begin{array}{cc}
\mathbf{1}_+ & 0  \\
0 & -\mathbf{1}_- 
\end{array}
\right)
,
\end{equation}
where $\mathbf{1}_\pm$ are identity matrices of the same size as $H_{\text{eff},\pm}$.
Finally, as the position operator $Y$ anticommutes with $M_y$, we can write it as 
\begin{equation}
Y = 
\left(
\begin{array}{cc}
0 & Y_-  \\
Y_+ & 0 
\end{array}
\right)
,
\end{equation}
where $Y_+$ and $Y_-$ are diagonal matrices constructed from the $y$ coordinates of the symmetric and antisymmetric reduced systems.

\begin{figure}
  \centering
  \includegraphics[width=1\textwidth]{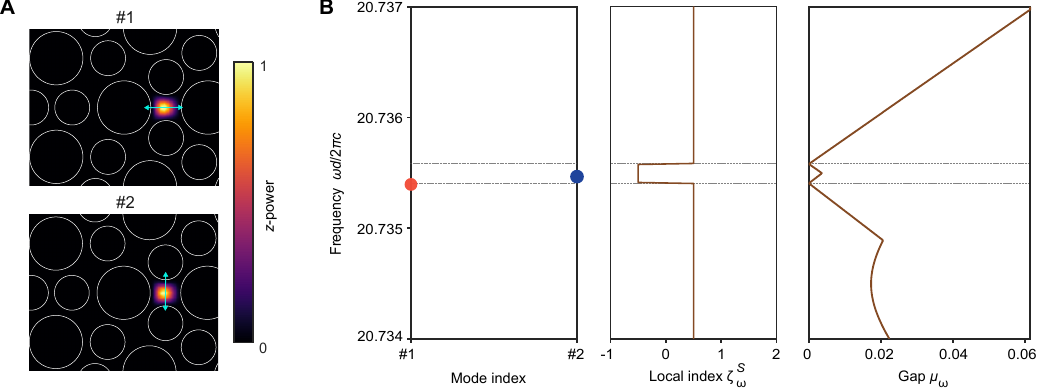}
  \caption{\textbf{Spectral localizer results for bent fiber.} (\textbf{A}) Calculated intensity profiles (normalized power flow in $z$) for a pair of radially and azimuthally polarized GTDMs. Polarization directions are indicated by cyan arrows. (\textbf{B}) Eigenmode spectrum (left panel), local index $\zeta_\omega^{S}$ (center panel) and local gap $\mu_\omega$ (right panel) from spectral localizer calculations.  In these calculations, we use $k_z d/2\pi = 30$ and bending radius $R=1.5$~$\textrm{cm}$.}
  \label{fig:extended_data_2}
\end{figure}

We calculate the spectral localizer for a structure based on the ideal TPCF preform hole profile (Fig.~\ref{fig:extended_data_0}E), which is mirror symmetric with respect to the $x$ axis.  To aid the analysis, a mirror-symmetric refractive index perturbation is introduced to break the degeneracy between the doublet states among the GTDMs; we opt for a perturbation consistent with fiber bending (Eq.~\eqref{eq:ri_bending}).  In Fig.~2G and Fig.~\ref{fig:extended_data_2}, we use a bending radius of $R = 1.5~\text{cm}$, along with $y = 0$ and $\kappa = \kappa_0 \left[ 10^{-4} \Vert H_{\text{eff},c}(\omega_0) \Vert / \Vert Y_c \Vert \right]$.  The local gap is normalized by $10^{-4} \Vert H_{\text{eff},c}(\omega_0) \Vert$, where $\Vert \cdots \Vert$ is the spectral norm (i.e., largest singular value).  Moreover, Fig.~2G uses $\kappa_0 = 0.01$ and $\omega_0 = 1.5 (2\pi c/d)$, while Fig.~\ref{fig:extended_data_2} uses $\kappa_0 = 0.1$ and $\omega_0 = 20.73 (2\pi c/d)$.

\begin{figure}
  \centering
   \includegraphics[width=1\textwidth]{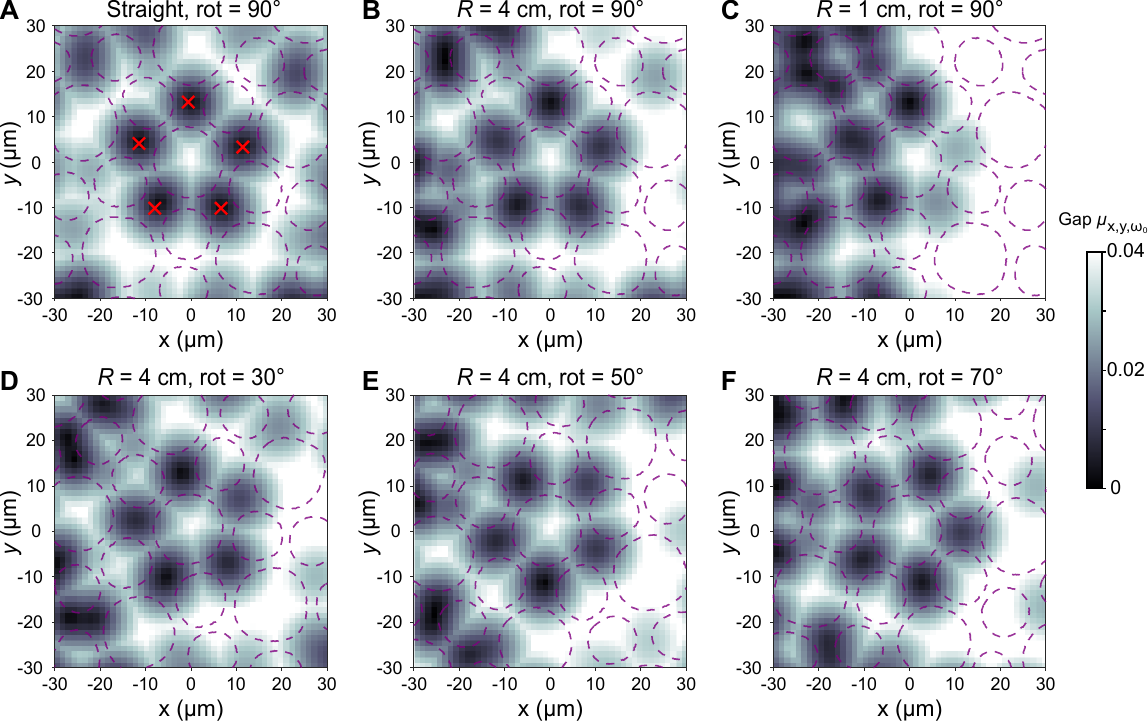}
   \caption{\textbf{Characterization of GTDM robustness by local gap.} (\textbf{A}) Local gap $\mu_{x,y,\omega_0}$ calculated at $k_z d/2\pi = 30$ and $\omega_0 = 20.72 (2\pi c/d)$ for the measured hole profile, consisting of the structure of Fig.~\ref{fig:extended_data_0}E rotated counter-clockwise by $90^\circ$.  The five zeros of $\mu_{x,y,\omega_0}$ near the disclination core (red crosses) correspond to the GTDMs, previously identified as topological states via jumps in a local index (Fig.~2G).  (\textbf{B} and \textbf{C}), Local gaps calculated at the same $k_z$ and $\omega_0$ for structures deformed by bending the fiber along $x$ with bending radius $R=4\,\textrm{cm}$ (B) and $R=1\,\textrm{cm}$ (C). (\textbf{D}---\textbf{F}), Local gaps calculated at bending radius of $R=4\,\textrm{cm}$, but from the measured hole profile rotated counter-clockwise by $30^\circ$ (D), $50^\circ$ (E) and $70^\circ$ (F). This bending spoils the mirror symmetry previously assumed in the symmetry-reduced spectral localizer analysis. }
  \label{fig:extended_data_3}
\end{figure}

\subsubsection*{S5.~2D localizer for bending analysis}
\label{secJ}

The spectral localizer can also quantify how robust the GTDMs are to perturbations that do \textit{not} obey the previously-assumed mirror symmetry.  To accomplish this, we employ the 2D localizer \cite{Garcia2024, Cerjan_JMP2023_NH} 
\begin{equation}
\label{eq:localizer_2d}
L_{x,y,\omega}(X_c,Y_c, H_\text{eff,c}) = 
\left(
\begin{array}{cc}
H_\text{eff,c}(\omega) & \kappa \left( X_c - x \mathbf{1}_c \right) - i \kappa \left( Y_c - y \mathbf{1}_c \right)  \\
\kappa \left( X_c - x \mathbf{1}_c \right) + i \kappa \left( Y_c - y \mathbf{1}_c \right) & -H_\text{eff,c}^\dagger(\omega) 
\end{array}
\right).
\end{equation} 

Unlike the 1D localizer in Eq.~\eqref{eq:reduced_localizer}, this can probe 2D positions $(x,y)$ without assuming the structure to be mirror symmetric or the system's Hamiltonian to be Hermitian.  We do not use this to formulate a local index, instead focusing on the local gap measure (also called a ``linear local gap'') \cite{Garcia2024},

\begin{equation}
\label{eq:local_gap_2d}
\mu_{x,y,\omega}(X_c, Y_c, H_\text{eff,c}) = \text{min} \left| \text{Re} \left( \text{spec} \left[ L_{x,y,\omega}(X_c, Y_c, H_\text{eff,c}) \right) \right] \right|,
\end{equation}

where $\text{spec}[\cdots]$ denotes the spectrum of the matrix.
The condition $\mu_{x,y,\omega} \approx 0$ is associated with the existence of a localized state near $(x,y)$ at frequency $\omega$. 

%% Any change of $\mu_{x,y,\omega}$ is bounded by the perturbations $\delta H_\text{eff,c}$ through Weyl's theorem \cite{Alex_APL2024}
%% % 
%% \begin{equation}
%% \label{eq:weyl_thm}
%% \left| \mu_{(x_1,y_1,\omega_1)}(X_c,Y_c,H_\text{eff,c}) - \mu_{(x_1,y_1,\omega_1)}(X_c,Y_c,H_\text{eff,c}+\delta H_\text{eff,c}) \right| \leq \Vert \delta H_\text{eff,c} \Vert
%% .
%% \end{equation}
%% %  
%% Therefore, for the topological modes to be localized at another position $(x_1, y_1, \omega_1)$ when the system is perturbed requires $\mu_{(x_1, y_1, \omega_1)}(X_c,Y_c,H_\text{eff,c}^\text{(pert)}) \rightarrow 0$ at that location.
%% In other words, for a topological mode to be moved, the system would need to be perturbed strong enough (irrespective of mirror symmetry-preserving perturbations) such that 
% 
%% \begin{equation}
%% \label{eq:pert_cond}
%% \Vert \delta H_\text{eff,c} \Vert \geq \mu_{(x_1, y_1, \omega_1)}(X_c,Y_c,H_\text{eff,c})
%% .
%% \end{equation} 

We can now determine whether these zeros remain at these positions, or move away, when the ideal fiber structure is perturbed (such as through bending). 

We first consider the measured hole profile (see Fig.~\ref{fig:extended_data_0}E) and show the corresponding $\mu_{x,y,\omega}$ in Fig.~\ref{fig:extended_data_3}A. We indeed observe zeros of $\mu_{x,y,\omega}$ at the locations of the GTDMs (here we take $k_z d/2\pi = 30$, for which the GTDMs are near-degenerate), confirming the robustness of fiber design against fabrication distortions \cite{Alex_APL2024}.
In addition to taking into account the measured hole profile, we plot in Fig.~\ref{fig:extended_data_3}B and C the local gap $\mu_{x,y,\omega}$ for the perturbed structures corresponding to bending the fiber in the $x$-axis, with bending radii of $4\,\textrm{cm}$ and $1\,\textrm{cm}$ respectively. The fiber structure is obtained by rotating the measured hole profile by $90^\circ$ (counter-clockwise direction). The bending along the $x$-axis is simulated using a conformal transformation on the refractive index (in both dielectric and air) given Eq.~\eqref{eq:ri_bending}. When the bending becomes too large, e.g. for a bending radii of $1\,\textrm{cm}$ in Fig.~\ref{fig:extended_data_3}C, the zeros of $\mu_{x,y,\omega}$ associated with the GTDMs starts to merges with each others or with the bulk modes, indicating the break down of the GTDMs.

Bendings along arbitrarily axis are also considered by first rotating the measured hole profile with different angles and then applying a bending along the $x$-axis. 
Figures~\ref{fig:extended_data_3}D and F show $\mu_{x,y,\omega}$ for the fiber rotated, respectively, by $30^\circ$, $50^\circ$ and $70^\circ$, all for a given bending radius of $4\,\textrm{cm}$.
Altogether, the zeros of $\mu_{x,y,\omega}$ associated with the GTDMs remain nearly unchanged around the fiber core, demonstrating the robustness of the GTDMs against fiber bending \cite{Alex_APL2024}.

% In Fig.~\ref{fig:extended_data_3}\textbf{b}--\textbf{c}, we show $\mu_{x,y,\omega}$ for deformed structures corresponding to bending the fiber in the $x$ axis, with bending radii of $4\,\textrm{cm}$ and $1.5\,\textrm{cm}$ respectively.  
% Unlike the earlier perturbation, these deformations do \textit{not} respect the mirror symmetry $M_y$.  
% We find that the zeros of $\mu_{x,y,\omega}$ corresponding to the GTDMs remain at almost the same positions near the fiber core, confirming the robustness of the GTDMs with respect to fiber bending \cite{Alex_APL2024}.

%%%%%%%%%%%%%%%% SUPPLEMENTARY REFERENCES %%%%%%%%%%%%%%%

% Do NOT include a reference list in the supplement.
% All references must be in a single list at the end of the main text.
% The copyeditors will ensure that the correct reference list appears with each version of the paper
% (print, HTML, PDF, mobile app, metadata for bibliographic databases etc.)
%\begin{thebibliography}{1}

\end{document}